\documentclass[final]{ias2}

\usepackage{graphicx} 
\usepackage{multirow}
\usepackage{array} 
\usepackage{color}

\usepackage{hyperref} 

\newcommand{\ket}[1]{\left|#1\right\rangle}
\newcommand{\bra}[1]{\left\langle#1\right|}

\newcommand{\mbf}[1]{\mathbf{#1}}

\begin{document} 

\markboth{}{Vaibhav Madhok, et. al.}
\title{Characterizing and Quantifying Quantum Chaos with Quantum Tomography}
\date{\today}
    


\author[sin]{Vaibhav Madhok} 
\email{vmadhok@gmail.com}
\author[ain]{Carlos A. Riofr\'io}
\email{criofrioa@gmail.com}
\author[yain]{Ivan H. Deutsch} 
\email{ideutsch@unm.edu}

\address[sin]{Department of Mathematics and Department of Zoology, University of British Columbia}
\address[ain]{Dahlem Center for Complex Quantum Systems, Freie Universit\"{a}t Berlin, 14195 Berlin, Germany}
\address[yain]{Department of Physics and Astronomy, University of New Mexico}

\begin{abstract}
We explore quantum signatures of classical chaos by studying the rate of information gain in quantum tomography.  
The tomographic record consists of a time series of expectation values of a Hermitian operator evolving under application of the Floquet operator of a quantum map that possesses (or lacks) time reversal symmetry. We find that the rate of information gain, and hence the fidelity of quantum state reconstruction, depends on the symmetry class of the quantum map involved. 
Moreover, we find an increase in information gain and hence higher reconstruction fidelities when the Floquet maps employed increase in chaoticity. We make predictions for the information gain and show that these results are well described by random matrix theory in the fully chaotic regime. 
We derive analytical expressions for bounds on information gain using random matrix theory for different class of maps and show that these bounds are realized by fully chaotic quantum systems.

\end{abstract}    


 
\maketitle


\section{Introduction: Classical and Quantum Chaos}

 Classical chaos is characterized by the sensitive dependence on initial conditions in a deterministic dynamical system \cite{s94}. In a conservative Hamiltonian system, this occurs for trajectories which do not settle down to fixed points, periodic orbits, or quasi-periodic orbits in the limit $t \rightarrow \infty$, where $t$ is the time of evolution of the trajectory \cite{s94, idam14}. Sensitive dependence on initial conditions means that nearby trajectories separate exponentially fast; the rate of separation is given by the Lyapunov exponent, $\lambda$, which 
characterizes the dynamics of the system. A conservative Hamiltonian system with $N$ degrees of freedom, with $N$ constants of motion, it is said to be integrable, and its dynamics is regular. When there are fewer than $N$ constants of motion, then the individual trajectories can explore the phase space in a complex manner and the system can exhibit chaos. 
   
It is not difficult to see that the above ``definition" of chaos fails in the quantum domain. A quantum state is not a point in the phase space but is described by a state vector. The time evolution of the state vector, due to the  Schr$\ddot{\rm{o}}$dinger's equation, is unitary. This means that the overlap of two state vectors undergoing evolution is \textit{constant} with time. Therefore, quantum systems, unlike their classically chaotic counterparts, do not show a sensitive dependence on initial conditions.
Furthermore, while classical chaos can lead to infinitely fine structures in the phase space, in quantum mechanics, Planck's constant, $\hbar$, sets the scale for such structures, according to Heisenberg's uncertainty principle.
 This is often stated as the key reason for the absence of chaos in the quantum domain. This, however, is not the complete story. An alternate description of classical mechanics, involving the evolution of classical probability densities, preserves the distance between two probability densities as a function of time \cite{koopman}.  Hence, the distance between two probability densities does not show exponential sensitivity even for classical mechanics.
        
 All this leads to two interesting questions: 
\begin{enumerate}
\item How does classically chaotic dynamics inform us about certain properties of quantum systems, e.g., the energy spectrum, nature of eigenstates, correlation functions, and more recently, entanglement and quantum discord. Alternatively, what features of quantum systems arise due to the fact that their classical description is chaotic?
\item  Since all systems are fundamentally quantum mechanical, how does classical chaos, with trajectories sensitive to initial conditions, arise out of the underlying quantum equations of motion?   
 \end{enumerate}
 These two questions are not unrelated. However, the first question deals mainly with finding the signatures of chaos by studying the properties of the quantum Hamiltonian, while the second concerns with the dynamical behaviour of quantum states and the emergence of classically chaotic behaviour in the macroscopic limit.    
 
  A central result of quantum chaos is its relationship to the theory of random matrices \cite{Haake}.  In the limit of large Hilbert space dimensions (small $\hbar$), for parameters such that the classical description of the dynamics shows global chaos, the eigenstates and eigenvalues of the quantum dynamics have the statistical properties of an ensemble of random matrices.  The appropriate ensemble depends on the properties of the quantum system under time-reversal \cite{Haake}. The ensemble of random matrices used to describe the Hamiltonians unrestricted by the time reversal symmetry is the Gaussian Unitary Ensemble (GUE). Similarly, the ensemble of random matrices used to describe the Hamiltonians having a time reversal symmetry are given by the Gaussian Orthogonal Ensemble (GOE).  The other class of random matrices typically studied are the random unitary matrices. They are employed for periodically driven systems, as models of the unitary ``Floquet" operators, $F$, describing the change of the quantum state during one cycle of the driving. Powers of the ``Floquet" operator, $F^{n}$, give us a stroboscopic description of the dynamics.
The ensemble of random unitaries are also known as the ``circular ensembles", originally introduced by Dyson \cite{d62}. 
  As was the case for random Hermitian matrices, time reversal symmetry arguments play a similar role in the choice of the appropriate ensemble of random unitaries employed to model the ``Floquet" operator
to study the properties of the chaotic system. 
Depending on whether the system has time reversal symmetry or not, the appropriate ensemble of random unitaries is called the Circular Orthogonal Ensemble (COE) or the Circular Unitary Ensemble (CUE) respectively.    
The eigenvectors of the COE and CUE have the same properties as that for the respective GOE and GUE, but the eigenvalues are distributed differently. 
The circular unitary ensemble (CUE) is just the ensemble of random unitary matrices picked from $U(n)$ according to the Haar measure. CUE eigenvalues lie on the unit circle in the complex plane, and hence the name.
 
 From this fact, an important quantum signature of chaos was obtained by Bohigas and collaborators \cite{Bohigas}, describing the spectral statistics of quantum Hamiltonians whose classical counterparts exhibit complete chaos using random matrix theory. Such signatures of quantum chaos have mainly focussed on the time-independent Schr$\ddot{\rm{o}}$dinger's equation and features like energy spectra and eigenstates. 

  Though quantum systems show no exponential separation under the evolution of a known unitary evolution, they do show a sensitivity to the parameters in the Hamiltonian \cite{per84}. Peres \cite{per84} showed that the evolution of a quantum state is altered when a small perturbation is added to the Hamiltonian. As time progresses, the overlap of the perturbed and unperturbed states gives an indication of the stability of quantum motion. It was shown that if a quantum system has a classically chaotic analog, this overlap has a very small value. On the other hand, if the classical analog is regular, the overlap remains appreciable.
  In another perspective, as seen in the work of Schack and Caves, quantum systems exhibit chaos when they are perturbed by the environment. They become hypersensitive to perturbations \cite{sc96}, as seen in the information-theoretic studies of the cost to maintain low entropy in the face of loss of information to the environment. This particular feature of quantum chaotic systems has several interesting consequences. For example, Shepelyansky has done extensive work on the issue of many-body quantum chaos in the quantum computer hardware and its effect on the accuracy of quantum computation \cite{Shepelyansky} in the absence of error correction. Recently, classical simulations of quantum dynamics have been connected to integrability and chaos \cite{pz07}.

It is imperative to mention the role played by quantum information theory in the above journey.
Quantum information science has added a whole new perspective to the study of quantum mechanics. This has resulted in a better understanding of quantum phenomena like entanglement and decoherence, and given us the tools to view certain quantum properties of physical systems as a resource. This has also enabled us to address the key questions in quantum chaos from a new perspective. As mentioned above, this has led to an information theoretic characterization of quantum chaos \cite{sc96} and explained the exploration of the behavior of chaotic quantum systems in the presence of environment induced decoherence \cite{Zurek/Paz} along with its connection to the quantum-to-classical transition. The study of quantum chaos from a quantum information perspective is also closely related to the theory and application of random quantum circuits \cite{e03}.  In the last two decades, quantum information theory has given us a new perspective in finding the fingerprints of chaos in quantum mechanics. The dynamical generation of entanglement and discord and information gain in tomography have been studied as signatures of classical chaos in the quantum world \cite{fnp, l01, Lakshminarayan/Bandyopadhyay2002, Lakshminarayan, Miller/Sarkar, Ghose, Wang, Lakshminarayan/Bandyopadhyay2004, Jacquod, tmd08, mgtg15}.

 In \cite{mrgd14}, it was shown that information gain about an initial quantum state in the process of quantum tomography is a metric to characterize and quantify quantum chaos. In this work, we review this new information-theoretic characterization of chaos and show how this procedure can be used to distinguish between symmetry classes of various quantum maps.
 
  Quantum tomography is the process of estimating an unknown quantum state from the statistics of measurements made on many copies of the state.
  In this work, we extend our efforts on information gain in quantum tomography to characterize the properties of the underlying dynamics. In particular, we give new analytical results for the information gain
  for different classes of quantum maps depending on their time reversal and parity symmetry properties. 
 The standard way to perform quantum tomography is to make projective measurements of an ``informationally complete" set of observables and repeat them many times. The statistics obtained are used to estimate the expectation values of the observables and hence the unknown initial state.

 Projective measurements pose a hurdle in exploring the connections between information gain in tomography and chaos due to large measurement back-action on the system. However, we overcome this by employing the protocol
for tomography via weak continuous measurement developed by Silberfarb et al. \cite{sjd05}.
In this protocol, the ensemble is collectively controlled and probed in a time dependent manner to obtain an ``informationally complete" continuous measurement record. We consider the case of a very weak measurement such that the back-action is negligible. This is possible when the uncertainty in any measurement outcome is small compared to the quantum uncertainty associated with the probe itself. 
 We accurately model all of the quantum dynamics occurring in the system, and then use the measurement time history to give us information about the initial quantum state.  The dynamics is ``informationally complete" if the time history contains information about an arbitrary initial condition. Our goal is to characterize and quantify the performance of tomography, when the dynamics driving the system are chaotic in the classical limit. We use this to draw connections 
 between the role played by regular and chaotic dynamics as well as the nature of symmetries of the dynamics in the tomography procedure. The work presented in this paper is intimately related to the protocols that have recently been implemented in the laboratory \cite{jessen}.
 
The remainder of this paper is organized as follows.  In Sec. $2$, we review the protocol for tomography  via weak continuous measurement developed by Silberfarb et al. \cite{sjd05} and Riofr\'io et al. \cite{rjd11}. In Section $3$,  we demonstrate how information gain while performing tomography is a quantum signature of classical chaos.   We perform numerical simulations of the reconstruction fidelity and its relationship to the degree of chaos in the dynamics that drive the system.  We also show how the fidelity obtained and the corresponding metrics to quantify information gain can be used to distinguish quantum maps belonging to different symmetry classes. We then explain these results in terms of the properties of random states in Hilbert space. Our results are discussed and summarized in Sec. $4$. 
 
\section{Tomography via Weak Continuous Measurement}
In this section, we review tomography via a continuous measurement protocol. 
Consider an ensemble of $N$, noninteracting, simultaneously prepared quantum systems in an identical, but unknown, state described by the density matrix 
$\rho_0$. Our goal is to determine $\rho_0$ by continuously measuring an observable $\mathcal{O}_0$. The ensemble is collectively controlled and probed in a time-dependent manner to obtain an ``informationally complete" continuous measurement record. In order to achieve informational completeness, when viewed in the Heisenberg picture, the set of measured observables should span an operator basis for $\rho_0$. For a Hilbert space of finite dimension $d$, and fixing the normalization of $\rho_0$, the set of Hermitian operators must form a basis of $su(d)$.
The measurement record is inverted to get an estimate of the unknown state. Laboratory realization of such a record is intimately tied to \textit{controllability}, i.e., designing the system evolution is such a way as to generate arbitrary unitary maps. While it is desirable to obtain an informationally complete measurement record, we shall see that we can obtain high fidelity in tomography in some cases even when this is not the case ~\cite{mrfd10}.

 In an idealized form, the probe performs a QND measurement that couples uniformly to the \textit{collective variable} across the ensemble and measures $\sum_{n=j}^ N \mathcal{O}_0^{(j)}$. For a strong QND measurement, quantum backaction will result in substantial entanglement among the particles. For a sufficiently weak measurement, the noise on the detector (e.g., shot noise of a laser probe) dominates the quantum fluctuation intrinsic to the measurement outcomes of the state (projection noise). In this case, we can neglect the backaction on the quantum state and the ensemble remains factorized. In order to obtain a measurement record that can be inverted to reconstruct an estimate of the initial state, one must drive the system by a carefully designed dynamical evolution that continually maps new information onto the measured observable. In order to do so, the system is manipulated by external fields. The Hamiltonian of the system, $H(t) = H[\phi_{i}(t)]$, is a functional of a set of time dependent control functions, $\phi_i(t)$, so that the dynamics produces an informationally complete measurement record $\mathcal{M}$. 
 
 Then we can write the measurement record obtained as

\begin{equation}
\label{m_record}
\mathcal{M}(t) = Tr(\mathcal{O}_0 \rho(t)) + \sigma W(t),
\end{equation}
amplified by the total number of copies ($N$ atoms in this case). Here $\sigma W(t)$ is a Gaussian-random variable with zero mean and  variance $\sigma^{2}$, which accounts for the noise on the detector.
   Since our goal is to estimate the initial state from the measurement record and the system dynamics, we will work in the Heisenberg picture.
Rewriting Eq. \ref{m_record} in the Heisenberg picture, we get 
\begin{equation}
\label{m_record_heisenberg}
\mathcal{M}(t) = Tr(\mathcal{O}(t) \rho_{0}) + \sigma W(t).
\end{equation}
We sample the measurement record at discreet times so that
\begin{equation}
\label{m_record_heisenberg_d}
\mathcal{M}_i = Tr(\mathcal{O}_i \rho_{0}) + \sigma W_i.
\end{equation}
    Thus, the problem of state estimation is reduced to a linear stochastic estimation problem.
    
     The goal is to determine $\rho_0$, given $\{\mathcal{M}_i\}$ for a well chosen $\{\mathcal{O}_i\}$, in the presence of noise $\{W_i\}$.
 We use a simple linear parametrization of the density matrix
 
\begin{equation}
\label{density}
\rho_0 = \frac{I}{d} + \sum_{\alpha =1}^ {d^2-1} r_{\alpha} E_{\alpha},
\end{equation}
where $d$ is the dimension of the Hilbert space, $r_{\alpha}$ are $d^2-1$ real numbers (the components of a generalized Bloch vector), and $\{E_{\alpha}\}$ is an orthonormal Hermitian basis of traceless operators. We can then write Eq. \ref{m_record_heisenberg_d} as
  \begin{equation}
\label{m_record_heisenberg_d_parameter}
\mathcal{M}_i = \sum_{\alpha =1}^ {d^2-1} r_{\alpha} Tr(\mathcal{O}_i  E_{\alpha}) + \sigma W_i, 
\end{equation}
or, in the matrix form as
 \begin{equation}
\label{m_record_heisenberg_d_parameter_matrix}
\textbf{M} = \tilde{\mathcal{O}}\textbf{r} + \sigma \textbf{W},
\end{equation}    
which in general is an overdetermined set of linear equations with $d^2-1$ unknowns $\textbf{r} = (r_1, ..., r_{d^2-1}) $.

The conditional probablity distribution for the random variable $\textbf{M}$, given the state $\textbf{r}$, is the Gaussian distribution
 \begin{equation}
\label{probability_M}
\mathcal{P}(\textbf{M}|r) \propto \text{exp} ( - \frac{1}{2 \sigma^2} (\textbf{M} - \tilde{\mathcal{O}}\textbf{r})^{T}(\textbf{M} - \tilde{\mathcal{O}}\textbf{r})).
\end{equation}    
We can use the fact that the argument of the exponent in Eq. \ref{probability_M} is a quadratic function of $\textbf{r}$ to write the likelihood function (ignoring any priors) as
\begin{equation}
\label{probability_ML}
\mathcal{P}(r|\textbf{M}) \propto \text{exp} ( - \frac{1}{2 \sigma^2} (\textbf{r} - \textbf{r}_{ML} )^{T} (\textbf{r} - \textbf{r}_{ML} )),
\end{equation}  
 a Gaussian function over the possible states $\textbf{r}$ centered around the most likely state, $\textbf{r}_{ML}$, with the covariance matrix given by $\textbf{C} = \sigma^{2} (\tilde{\mathcal{O}}^{T}\tilde{\mathcal{O}})^{-1}$. The uncontrained maximum liklihood solution is given by
 \begin{equation}
\label{ML}
\textbf{r}_{ML} = (\tilde{\mathcal{O}}^{T}\tilde{\mathcal{O}})^{-1} \tilde{\mathcal{O}}^{T} \textbf{M}.
\end{equation}  
  The measurement record is “informationally complete” when the covariance matrix has full rank, $d^{2}-1$.
 If the measurement record is incomplete and the covariance matrix is not full rank,  we replace the inverse
 in Eq. \ref{ML} with the Moore-Penrose pseudo inverse \cite{ig03}.  The eigenvectors of $\textbf{C}^{-1}$ represent the orthogonal directions in operator space that we have measured up to the final time, and the eigenvalues  determine the uncertainty, or the signal-to-noise ratio, associated with those measurement directions.  
 
 When we have an incomplete measurement record, or in the presence of noise, the unconstrained maximum likelihood 
 procedure does not give a density matrix that corresponds to a physical state. The estimated density matrix might have negative eigenvalues. We correct this by finding a valid density matrix that is ``closest" to $\rho_{ML}$, the density matrix obtained by the unconstrained maximum likelihood procedure. 


\section{Information Gain in Tomography}
\subsection{Metrics to quantify information gain}

Our protocol for quantum tomography via continuous measurement of a driven system \cite{sjd05} gives us a window into the complexity of quantum dynamics and its relationship to chaos. Moreover, the experimental implementation of tomography by continuous measurement provides a useful platform for exploring these ideas in the laboratory \cite{jessen}. Quantum tomography deals with the extraction of information about an unknown quantum state through
measurements. In our attempt to study chaos under this paradigm, we define metrics to quantify this information gain. These metrics characterize the ability of our control dynamics to generate a sufficiently high signal-to-noise ratio for measurements in different directions of the operator space. As we shall see, these metrics elucidate the connection between the degree of chaos and the fidelities obtained in tomography.
We can quantify the information gain in a number of ways. 

1) \textbf{Fidelity of Tomography}: Fidelity of the reconstruction obtained in tomography is a metric for information gain which determines the degree of closeness of quantum states and is intimately related to how much information is obtained during the process. The fidelity is simply given by the overlap of the initial and the reconstructed state vectors. The fidelity between a target pure state $|\psi \rangle$ and the reconstructed state $\rho$ is $F = \bra{\psi}\rho\ket{\psi}$.
 
2) \textbf{Fisher Information (FI) of the Measurement Record:}
We can further quantify the correlation between chaos and the performance of quantum state estimation using information-theoretic metrics. The information obtained in measurement of a quantum system can be expressed in terms of the uncertainty of the outcomes summed over a set of mutually complementary experiments~\cite{brukner99} .  In terms of the Hilbert-Schmidt distance between the true and estimated state in quantum state reconstruction, averaged over many runs of the estimator, this information can be written as $I =\langle Tr\{(\rho_0 - \bar{\rho})^2\}\rangle $~\cite{rehacek02}, which in terms of the total uncertainty in the Bloch vector components is $I = \sum_\alpha\langle (\Delta r_\alpha)^2 \rangle$.  The Cramer-Rao bound tells us that this uncertainty obeys  
\begin{equation}
 \langle (\Delta r_\alpha)^2 \rangle \ge \left[ \mbf{F}^{-1} \right]_{\alpha \alpha},
\end{equation}
  where $\mbf{F}$ is the Fisher information matrix associated with the conditional probability distribution, Eq. \ref{probability_M}, and thus $I \ge Tr\, (\mbf{F}^{-1})$.

In general, for a multivariate parameter estimation problem, the Cramer-Rao bound gives
\begin{equation}
\label{Cramer-rao}
\textbf{Cov}_{\theta}{(T(X)}) \geq \textbf{F}^{-1},
\end{equation}
 where the matrix inequality, $A \geq B$, is understood to mean that the matrix, $A - B$, is positive semidefinite. Here $X$ is a $d$-dimensional random vector that contains information about the multivariate parameter,  $\theta = [{\theta_{1}, \theta_{2}, ..., \theta_{d}}]$, $\textbf{T(X)}$  is the unbiased estimator of the multivariate parameter,  and $\textbf{Cov}_{\theta}(T(X)) $ is the covariance matrix of a set of unbiased estimators for the parameters $\theta$. It quantifies the error in our estimation process. $\textbf{F} $ is the multivariate generalization of the \textbf{FI} \cite{ct},
\begin{equation}
\label{Multi-fisher}
F_{mn} = \textbf{E}(\frac{\partial}{\partial \theta_{m}} \text{log} f(x;\theta) \frac{\partial}{\partial \theta_{n}} \text{log} f(x; \theta)),
\end{equation}
where $f(x;\theta)$ is the probability density of the random variable $X$ conditioned on the value of $\theta$, and $E$ denotes the expectation value.
  In the limit of negligible quantum backaction, we saturate this bound. This is because our probability distribution is Gaussian, regardless of the state.  In that case, the Fisher information matrix equals the inverse of the covariance matrix, $\mbf{F} = \mbf{C}^{-1}$, in units of $\mbf{N^2/\sigma^2}$
and thus the Cramer-Rao bound reads
\begin{equation}
\label{CR-gaussian}
\textbf{Cov}_{\theta}(T(X)) \geq \textbf{C}.
\end{equation}
We consider the basis in which $F$, and hence $\textbf{C}^{-1}$, is diagonal,
\begin{equation}
\label{Fisher-diagonal}
F^{'} = U F U^{T}.
\end{equation}
Such a transformation is provided by $U$ composed from the eigenvectors of $\textbf{C}$. In this representation, the estimate of the newly transformed parameters fluctuate independently of each other. This suggests the possibility to form a single number that quantifies the performance of the tomography scheme as a whole by adding those independent errors, $\epsilon$, as
\begin{equation}
\label{error}
\epsilon \geq Tr(\textbf{C}).
\end{equation}
   Thus, $\frac{1}{Tr(C)}$, which is the collective FI, serves as a measure of the amount of information about the parameter $\theta$ that is present in the data. 

3) \textbf{Shannon Entropy of Eigenvalues of the Inverse Covariance Matrix}:
 The mutual information, $\mathcal{I}[\mbf{r};\mbf{M}]$, quantifies the information we have about parameters $\mbf{r}$ from measurement record $\mbf{M}$, which is given by $\mathcal{I}[\mbf{r};\mbf{M}] = H(\mbf{M}) - H(\mbf{M}|\mbf{r})$~ \cite{ct}.  Here $H$ is the Shannon entropy of the given probability distribution.   The entropy of the measurement record, $H(\mbf{M})$, arises solely due to the shot noise in the probe, and hence is a constant.  
The mutual information between the Bloch vector and a given measurement record can be expressed as the entropy of the conditional probability distribution (Eq. \ref{probability_M}) 
\begin{equation}
\mathcal{I}[\mbf{r};\mbf{M}] = -H(\mbf{M}|\mbf{r}) = - \frac{1}{2}\log\left(\text{det}\mbf{C}\right) = \log (1/V),
\end{equation}
where $V$ is the volume of the error-ellipsoid whose semi-major axes are defined by the covariance matrix.


\subsection{The Quantum Kicked Top}
\label{FQT}
How does the presence of chaos in the control dynamics influence our ability to perform tomography? In order to address this question, we chose the ``kicked top" dynamics \cite{Haake} as the paradigm to explore quantum chaos in tomography.
The Hamiltonian for the kicked top (after setting $\hbar$ = 1) is given by 

  \begin{equation}
\label{kicked_top}
H(t) = \frac{1}{\tau}p  J_{x} + \frac{1}{2j} \kappa  J_{z}^{2} \sum_{n=-\infty}^\infty \delta (t - n \tau).
\end{equation} 
Here, the operators, $J_{x}$, $J_{y}$ and $J_{z}$ are the angular momentum operators obeying the commutation relation $[J_{i}, J_{j}] = i \epsilon_{ijk} J_{k}$
The first term in the Hamiltonian describes a precession around the \textit{x} axis with an angular frequency $\frac{p}{\tau}$, and the second term describes a periodic sequence of kicks separated by time period $\tau$. Each kick is an impulsive rotation about the \textit{z} axis by an amount proportional to $J_z$.
Choosing the external field to act in delta kicks allows us to express the Floquet map (transformation after one period) in a simple form of sequential rotations as
\begin{equation}
\label{Floquet}
U_{\tau} = e^{\frac{-i \lambda J_z^2}{2j}} e^{-i\alpha J_{x}},     
\end{equation}
where $\alpha$ and $\lambda$ are related to $p$ and $\kappa$, respectively, in terms of the kicking period.
 The evolution of the initial quantum state has the form $U^n\rho{U^{\dagger n}}$, where $n$ enumerates the kick number or the periodic application of the map.
The classical map can be obtained by considering the Heisenberg evolution of the expectation values of the angular momentum operators in a familiar way \cite{Haake}. 
 
The classical dynamics consists of the motion of a unit spin vector
on the surface of the sphere.  The $z$-component of a spin and the angle $\phi$, denoting its orientation in the $x$-$y$ plane, are canonically conjugate, and thus the spin constitutes one canonical degree of freedom.  The classical dynamical map has the same physical action as described above in the quantum context -- precession of the spin around the \textit{x} axis with an angular frequency $\alpha$ followed by  an impulsive rotation around the \textit{z} axis by an amount proportional to $J_{z}$ with a proportionality constant $\lambda$.
In our analysis, we fix $\alpha = 1.4$ and choose $\lambda$ to be our chaoticity parameter. As we vary $\lambda$ from $0$ to $7$, the dynamics change from highly regular to completely chaotic.  
Since the total magnitude of the spin is a constant of motion, our classical map is two dimensional.  We visualize the phase by plotting the \textit{z} and \textit{y} components of motion after every application of the dynamical map. 

\begin{figure}
\includegraphics[width=\linewidth]{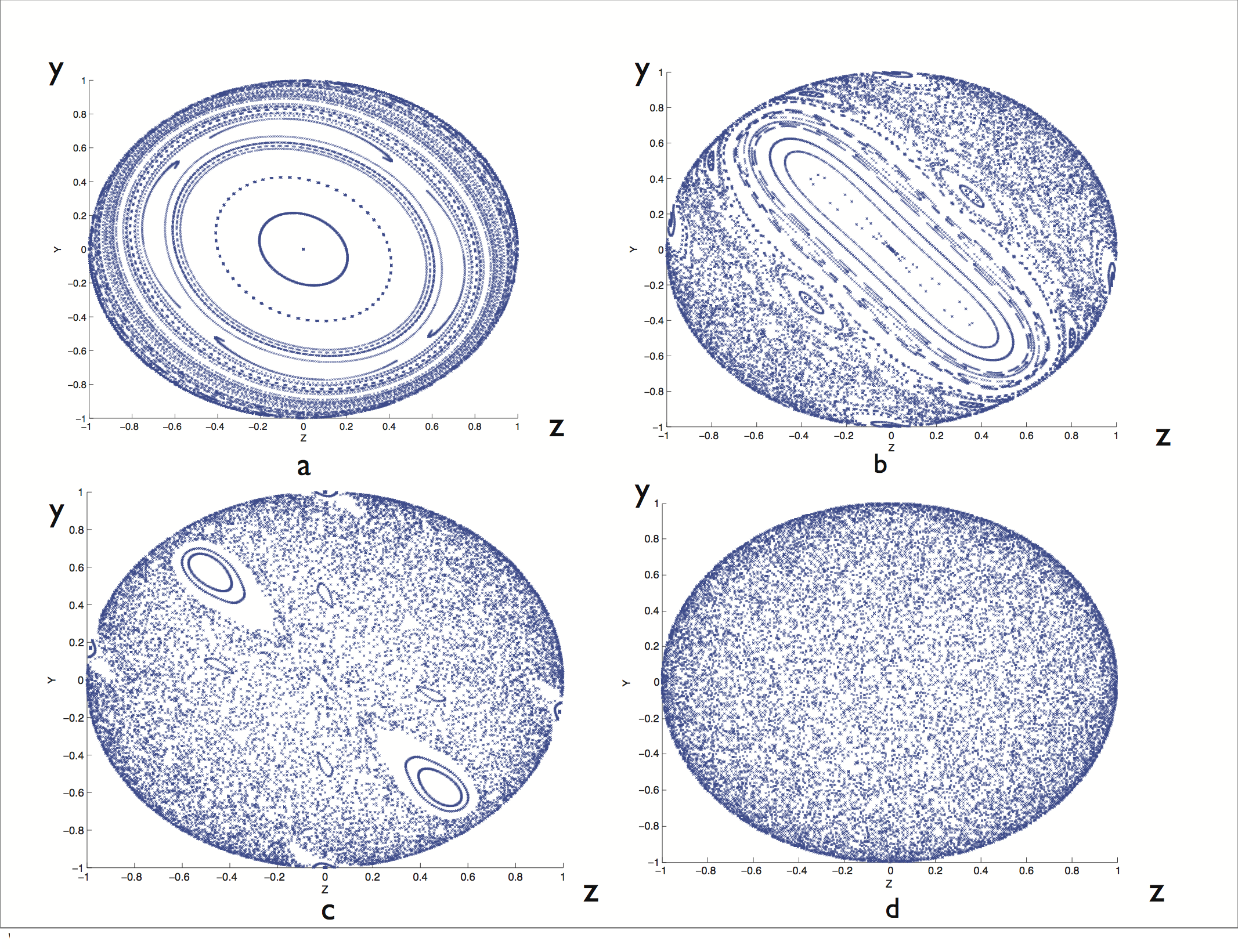}
\caption{Phase space plots for the kicked top in four regimes. (a) Regular phase space: $\alpha=1.4,\lambda=0.5$, (b) Mixed phase space: $\alpha=1.4,\lambda=2.5$, (c) Mostly chaotic: $\alpha=1.4,\lambda=3.0$, (d) Fully chaotic phase space: $\alpha=1.4,\lambda=7.0$. The figures depict trajectories on the southern hemisphere ($\textit{x} < 0$) of the unit sphere where $X = \frac{J_x}{j}$, $Y = \frac{J_y}{j}$ and $Z = \frac{J_z}{j}$, and we take the limit $j \rightarrow \infty $ to get the classical limit as in \cite{Haake}.}
\label{FKT}
\end{figure}

Figure \ref{FKT} shows four different regimes of classical dynamics.  With the parameters $\alpha=1.4, \lambda= 0.5$ (Fig. \ref{FKT}a), the dynamics are highly regular.  When $\alpha=1.4$, and $\lambda=2.5$ (Fig. \ref{FKT}b), we see a mixed space with chaotic and regular regions of comparable size. The parameters, $\alpha=1.4, \lambda = 3.0$ (Fig. \ref{FKT}c), give a phase space that has mostly chaotic regions and finally, $\alpha = 1.4, \lambda = 7.0$ gives a completely chaotic phase space (Fig. \ref{FKT}d).

A central result of quantum chaos is the connection with the theory of random matrices \cite{Haake}.  In the limit of large Hilbert space dimensions (small $\hbar$), for parameters such that the classical description of the dynamics shows global chaos, the eigenstates and eigenvalues of the quantum dynamics have the statistical properties of an ensemble of random matrices \cite{Bohigas}.  The appropriate ensemble depends on the properties of the quantum system under time-reversal symmetry\cite{Haake}.  We thus seek to determine whether there exists an anti-unitary (time reversal) operator $T$ that has the following action on the Floquet operator,
\begin{equation}
T U_{\tau} T^{-1} = U_{\tau}^{\dagger} = e^{i\alpha J_{x}}e^{\frac{i \lambda J_z^2}{2j}}.
\end{equation}
 Considering the generalized time reversal operation
\begin{equation}
\label{Eq:Reversal}
T=e^{i \alpha J_x} K,
\end{equation}
where $K$ is the complex conjugation operator.  
It then follows that 
\begin{eqnarray}
T U_{\tau} T^{-1} &=& \left( e^{i \alpha J_x} K \right) \left( e^{\frac{-i \lambda J_z^2}{2j}} e^{-i\alpha J_{x}} \right) \left( K e^{-i \alpha J_x} \right)  \\
&=&  e^{i \alpha J_x} \left( e^{\frac{+i \lambda J_z^2}{2j}} e^{i \alpha J_x}\right) e^{-i \alpha J_x} \nonumber\\
&=& e^{i \alpha J_x} e^{\frac{+i \lambda J_z^2}{2j}} = U_{\tau}^{\dagger}, \nonumber
\end{eqnarray}
so the dynamics is time-reversal invariant.  Moreover as $T^2=1$, there is no Kramer's degeneracy. Given these facts, for parameters in which the classical dynamics is globally chaotic, we expect the Floquet operator to have the statistical properties of a random matrix chosen from the circular orthogonal ensemble (COE) \cite{Haake}.

In order to have maximum information gain, we need to condition the dynamics so that we maximize $1/V = \sqrt{\text{det }(\mbf{ C^{-1}})}$. The quantity $Tr(\mbf{C}^{-1})$ is constrained at $t_n$. One can show that after $n$ steps
\begin{equation}
\label{eq:TraceInvC}
Tr(\mbf{C}^{-1}) = \sum_{i,\alpha} \left( \mathcal{O}_{i,\alpha}\right)^2 = n \| \mathcal{O} (0)\|^2,
\end{equation}
where $\| \mathcal{O}(0) \|^2=\sum_\alpha Tr (\mathcal{O}(0) E_{\alpha})^2$ is the Hilbert-Schmidt square norm, and  $\mathcal{O}(0) =J_z$ for our case. Therefore, from the theorem of the arithmetic and geometric means,
 \begin{equation}
 \text{det }(\mbf{ C^{-1}}) \le \left(\frac{1}{D} Tr(\mbf{C}^{-1}) \right)^D = \left(\frac{n}{D}\| \mathcal{O} (0)\|^{2}\right)^D,
 \end{equation}
where $D =d^2-1$ is the rank of the regularized covariance matrix. The maximum possible value of the mutual information is attained when all eigenvalues are equal, saturating the above inequality. At a given time step, the dynamics that gives the largest mutual information is the one that makes the eigenvalues most equal.  
If we normalize the eigenvalues of the inverse of the covariance matrix, then as a probability distribution, its Shannon entropy $E$, is a measure of how evenly we have sampled all the directions in the operator space.
 We reach the maximum entropy when we have measured all directions in the space of matrices equally, $E_{max}  = \text{log} (d^2 -1)$. This is the most unbiased measurement we can implement that will lead to the highest fidelities, on average, for a random state.    


\begin{figure}
\includegraphics[width=\linewidth]{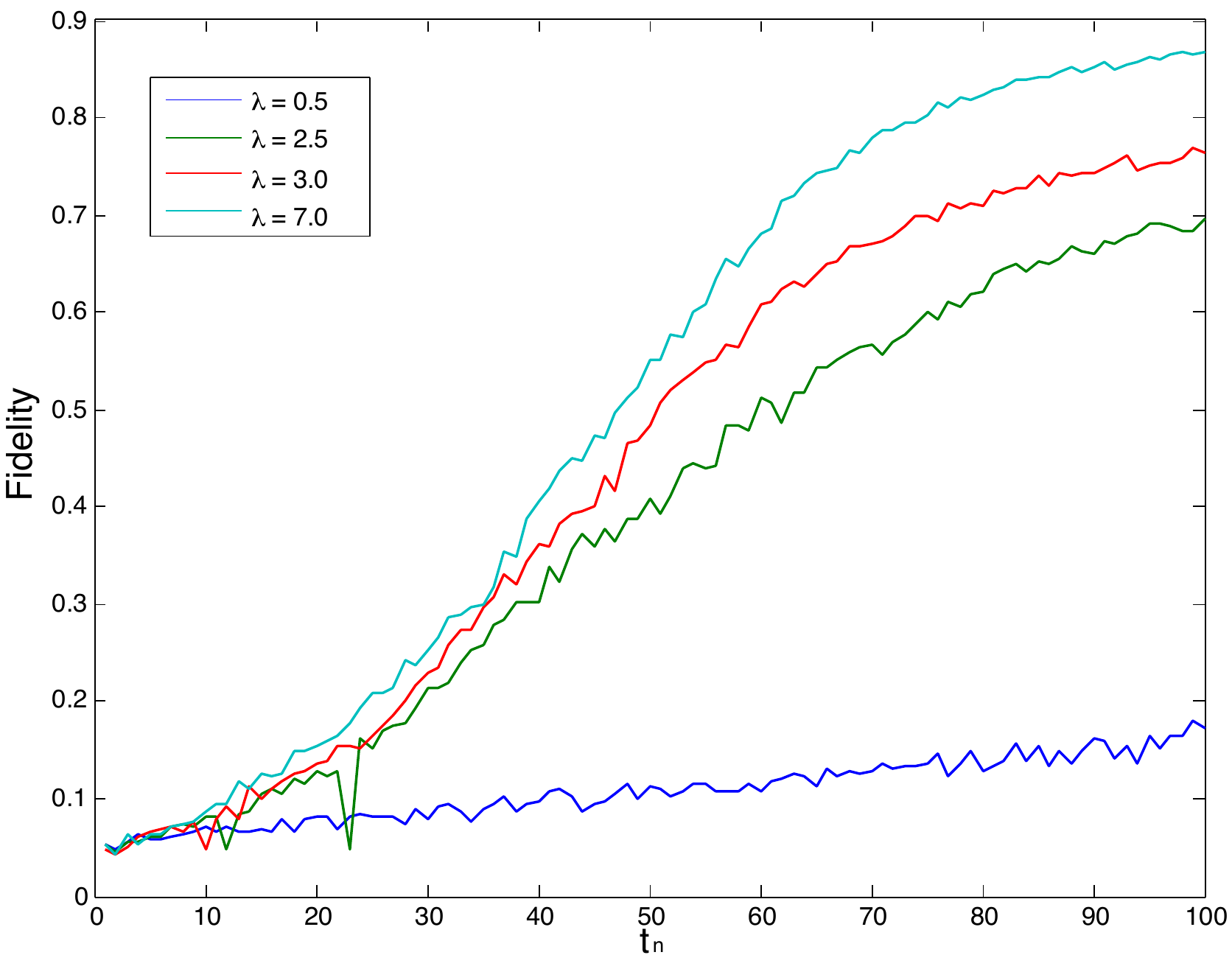}
\caption{Fidelity of reconstruction as a function of the number of applications of the kicked top map. The fidelity is calculated as the average fidelity of reconstruction of 100 states picked at random according to the Haar measure. The parameters of the kicked top are as described in the text, with $\alpha = 1.4$ fixed. We show the fidelity for different choices of the chaoticity parameter. Both the rate of growth and the final value of the fidelities are increased with higher values of $\lambda$.}
\label{F2}
\end{figure}

 \begin{figure}
\includegraphics[width=\linewidth]{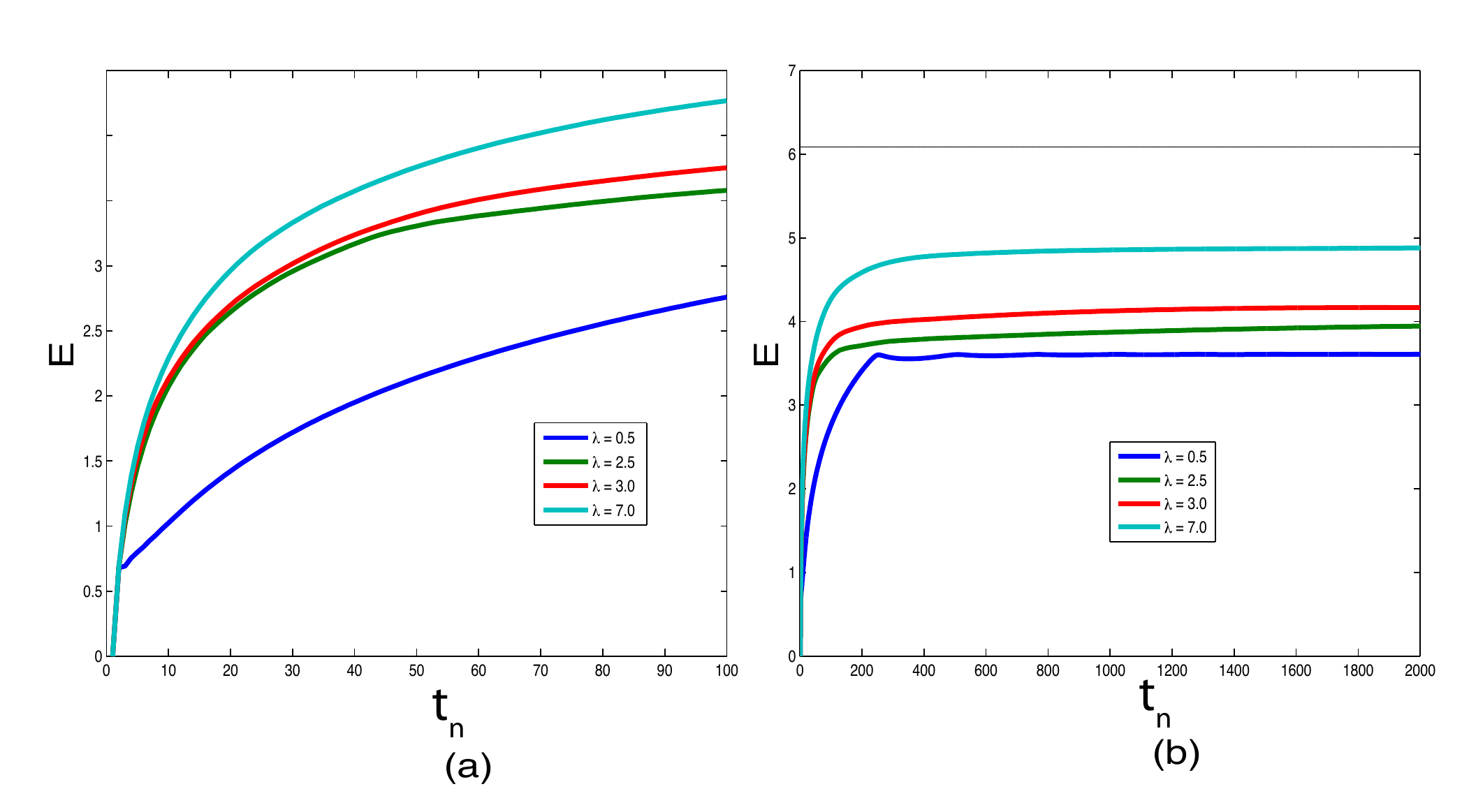}
\caption{The Shannon entropy of the normalized eigenvalues of the inverse of covariance matrix as a function of the number of applications of the kicked top map: (a) Short time behavior (b) long time/asymptotic behavior. The parameters are as described in the text.}
\label{F3}
\end{figure}
\begin{figure}
\includegraphics[width=\linewidth]{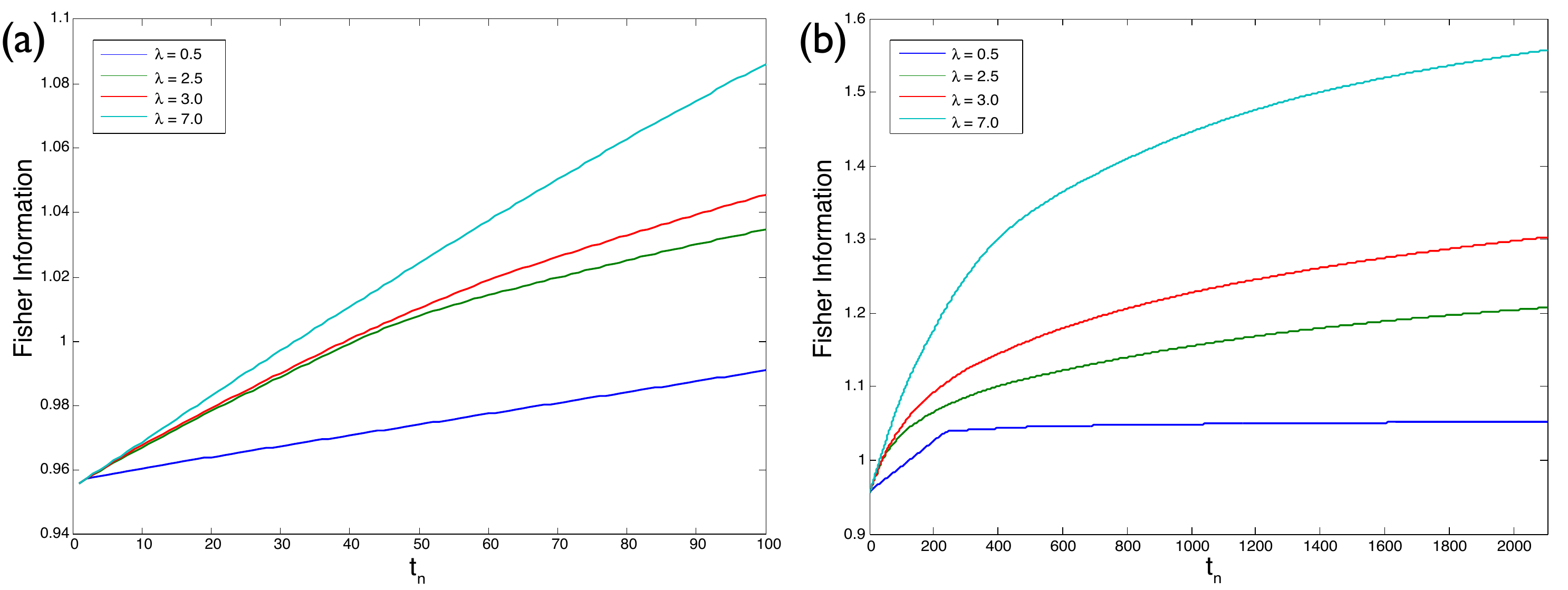}
\caption{The FI of the parameter estimation in tomography as a function of the number of applications of the kicked top map: (a) Short time behavior (b) long time/asymptotic behavior. The parameters are as described in the text.}
\label{F4}
\end{figure}

 The collective FI, $\frac{1}{Tr(\textbf{C})}$, tells us about the amount of information our measurement record contains about the parameters that define the density matrix. Figure \ref{F4} shows the behavior of the FI as a function of the number of applications of the kicked top map, and for different values of the chaoticity parameter. We see that the rate of increase of the FI is correlated with the degree of chaos present in the control dynamics. As our dynamics become increasingly chaotic, we obtain higher values for the FI at a given time. We expect the FI to be correlated with the average fidelities of estimation for an ensemble of random states.
\subsection{Signatures of chaos : Information gain in the fully chaotic regime and random matrix theory}
\subsubsection{Results and Discussion}
\label{RMTT}
We are now ready to explore the role of chaos in the performance of tomography. Throughout this section, we consider spin $J=10$, a $d=21$ dimensional Hilbert space, which is sufficiently large that a minimum uncertainty spin coherent state is a sufficiently confined ``wavepacket" that it can resolve features in the classical phase space.
Figure \ref{F2} shows the average fidelity of reconstruction of 100 states picked at random according to the Haar measure as a function of the number of applications of the kicked top map, and for different values of the chaoticity parameter. We see that the rate of increase in fidelity increases with the degree of chaos. The final fidelity achieved after a fixed number of kicks is also correlated with the degree of chaos. 
We can understand the above results by studying the information gain in tomography as a function of the degree of chaos in the control dynamics. Figure \ref{F3} shows the behavior of the entropy $E$ of the covariance matrix, as defined above, as a function of the number of applications of the kicked top map, and for different values of the chaoticity parameter. We see that the rate of increase of entropy for short times, Fig. \ref{F3}a, is correlated with the degree of chaos present in the control dynamics. The asymptotic value of the entropy reached also increases with the chaoticity parameter. Chaotic dynamics provides a measurement record with a large signal-to-noise ratio in all the directions in the operator space.
An increase in the chaoticity parameter results in an increasingly unbiased measurement process that will yield 
high fidelities for estimating random quantum states. Figure \ref{F3}a shows the behavior of the entropy at short time scales, while we see asymptotic behavior in Fig. \ref{F3}b.

\begin{figure}
\includegraphics[width=\linewidth]{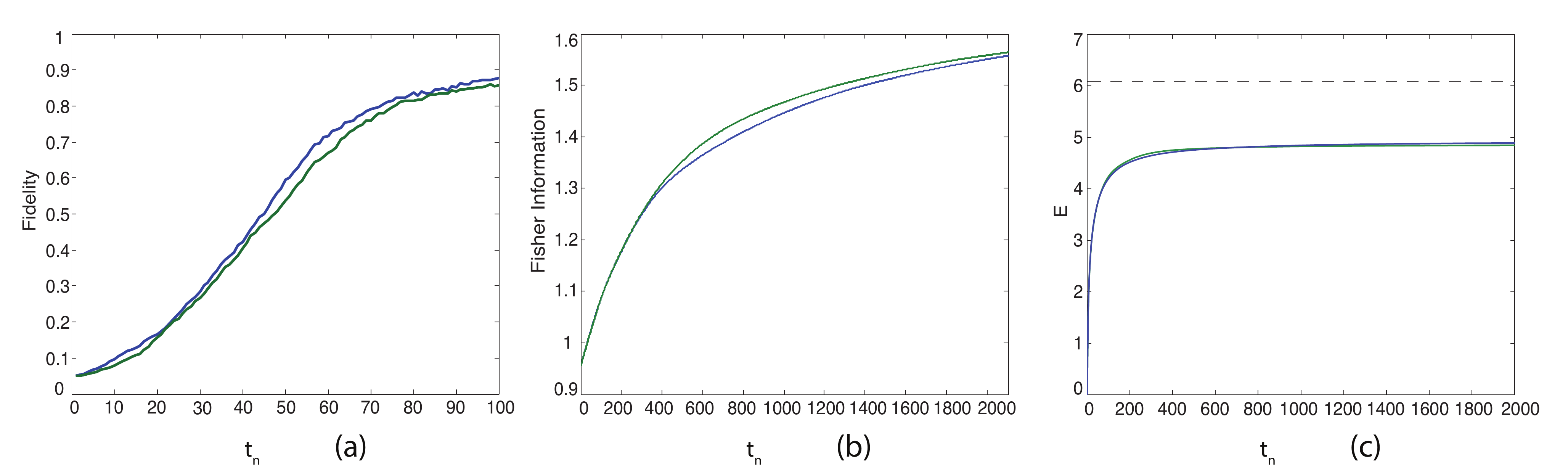} 
\caption{Comparison between the tomography performed by the repeated application of kicked top in the fully chaotic regime (the blue line) and that by a typical random unitary picked from the COE (the green line). (a) The average fidelity of reconstruction of 100 states picked at random according to the Haar measure. (b) The Shannon entropy of the normalized eigenvalues of the inverse of covariance matrix as a function of the number of applications of the map.  The dotted line gives the upper bound on the entropy, $E_{max} = \log (d^2 -1 )$.}
\label{F5}
\end{figure}

When the system is driven by dynamics that are completely chaotic, we expect the information gain and the fidelity to follow the predictions from random matrix theory.
Figure \ref{F5} shows the behavior of the fidelity, Shannon entropy and the FI of the inverse of the covariance matrix as a function of the number of applications of the kicked top map (the blue line) and compares them with the corresponding quantities for a typical random unitary picked from the COE (the green line). We see a strong agreement between our predictions from random matrix theory and the entropy calculation for the evolution by a completely chaotic map.
 
\begin{figure}[!htp]
\includegraphics[width = \linewidth]{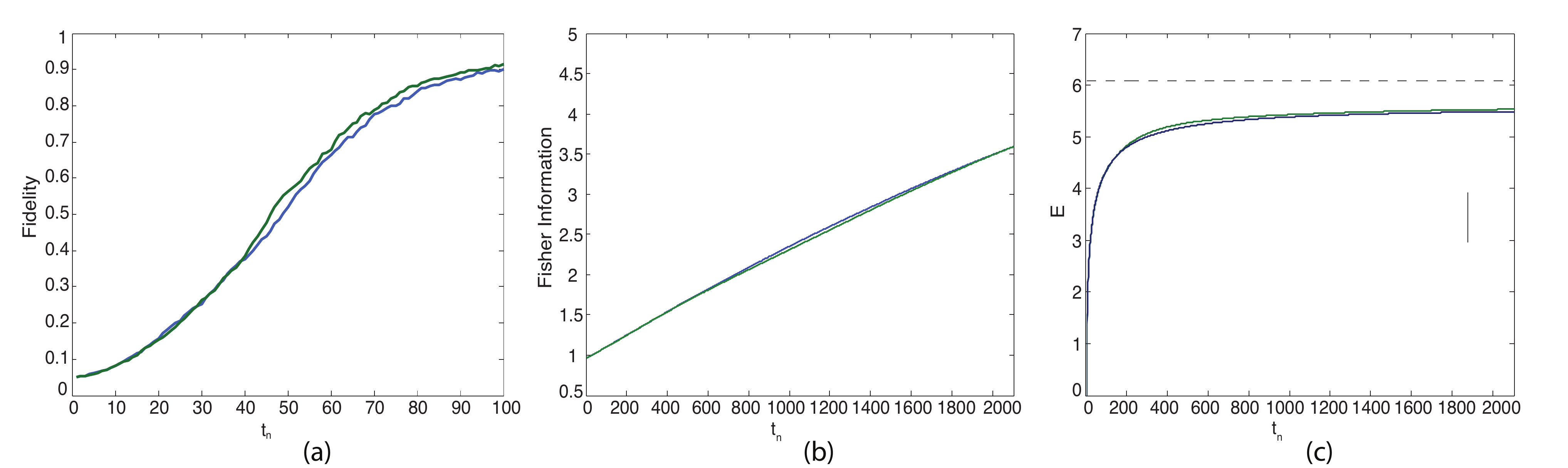}
\caption{Same as Figure 4, but for a kicked top without time reversal invariance (Eq. \ref{Floquet_noTR}) (the blue line). In this case as well, the results are well predicted by modeling the dynamics by random matrices sampled from the CUE (the green line).}
\label{F7}
\end{figure}

We test our predictions from  the random matrix theory for chaotic maps without a time reversal symmetry.
For example, another type of the ``kicked top" map without time reversal symmetry~\cite{kmh88} is given by
\begin{equation}
\label{Floquet_noTR}
U_{\tau}= e^{-i \lambda_1 - i J_x^2-i\alpha_1 J_{x}} e^{-i \lambda_2 - i  J_y^2-i\alpha_2 J_y} e^{-i \lambda_3 J_z^2-i\alpha_3 J_{z}}.
\end{equation}
In Fig. \ref{F7}, we repeat the above calculations for this map. In this case, the appropriate random matrix ensemble is the CUE. We see an excellent agreement between the behavior of the fidelity,  Shannon entropy and the FI, as predicted by random matrix theory, and that for the evolution by a completely chaotic map without the time reversal symmetry \cite{kmh88}.

When all the eigenvalues of the inverse of the covariance matrix are equal, we have an upper bound on the entropy,
 $E_{max}  = \text{log} (d^2 -1) $. Figures \ref{F5} - \ref{F7} compare the entropy values achieved 
 by the repeated application of the same unitary (time reversal invariant or otherwise) to $E_{max}$. We see that we fall significantly short of $E_{max}$ by such a procedure. 
 
\begin{figure}[!htp]
\includegraphics[width=\linewidth]{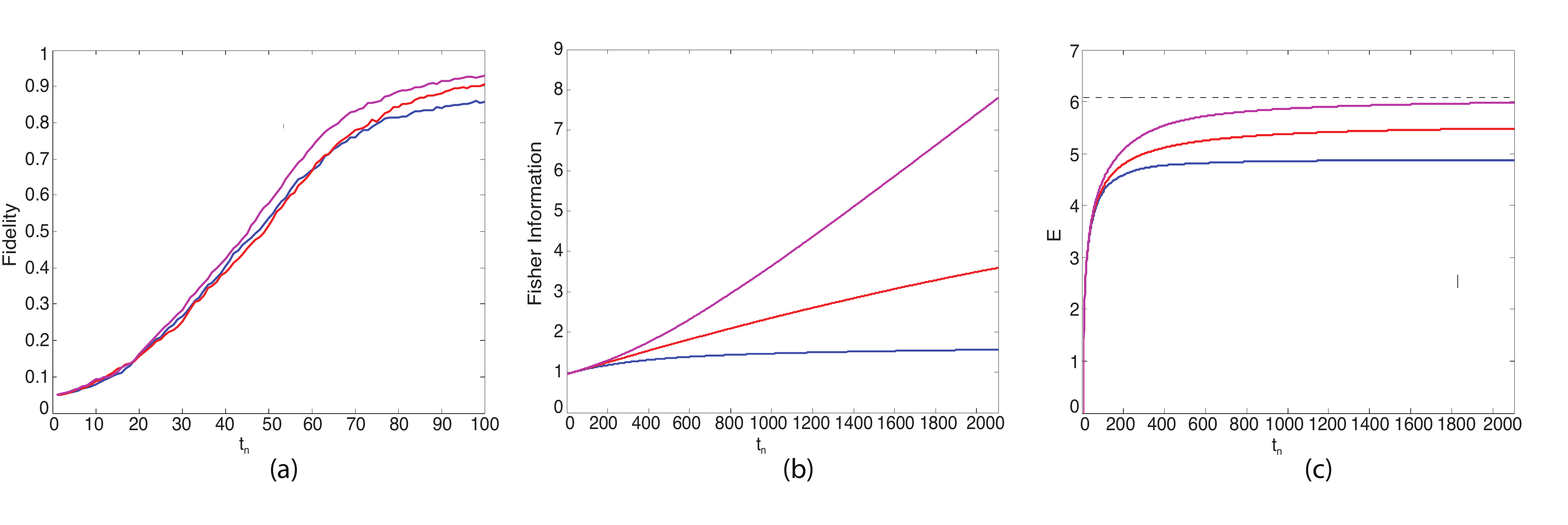}
\caption{Comparison between tomography performed by applying a \textit{different} random unitary at each time step, picked from the unitarily invariant Haar measure (magenta line) and that by a repeated application of a random unitary picked from the COE (blue line) and the CUE (red line) . The dotted line gives the upper bound on the entropy, $E_{max} = \log (d^2 -1 )$}
\label{F9}
\end{figure}

 So far, we have considered the application of the same unitary matrix periodically to obtain the measurement record.  However, this alone does not give us an informationally complete measurement record; high fidelities are reached only when we make use of the positivity constraint.  On the other hand, we can consider application of a series of \textit{different} random unitaries \cite{kkd15}.  In that case, we expect to rapidly reach an informationally complete set and thus rapidly gain information about tomography. In Fig. \ref{F9}, we plot the fidelities, Shannon entropy and the FI achieved by applying a \textit{different} random unitary, picked from the unitarily invariant Haar measure, and compare it with the results obtained by the repeated application of the same unitary (picked from the COE and CUE). We also see that we reach the upper bound, $E_{max}$, asymptotically, by this method. Indeed, an application of a different random unitary is the most unbiased dynamics we can hope to perform.   
                               
\subsubsection{Analytical expressions for Information gain}
In this section, we use random matrix theory to predict the information gain in tomography when we apply  the unitary map $U_\tau$ periodically.

 1) \textbf{The quantum kicked top} : 
In our system, we have a symmetry given by the parity operator $R$ that has the form
\begin{equation}
R = e^{-i\pi J_x}.
\end{equation}
Our unitary map, $U$, being the kicked top or the appropriate COE sampled matrix, will commute with $R$, i.e., $[R,U] = 0$. Thus, there exists a basis in which $U$ and $R$ are diagonal. First, note that the eigenvalues of $R$ are $\pm 1$. Then, let's define a basis, $\{|R_j\rangle\}$, where $|R_j\rangle=|R_j^{(-)}\rangle$ for $j = 1,\ldots,a$, and $|R_j\rangle=|R_j^{(+)}\rangle$ for $k=a+1,\ldots,d$, corresponding to the eigenvalues $-1$ and $+1$, and where $a\in\{(d+1)/2,(d-1)/2\}$. Since $U$ is also diagonal in this basis, an asymptotic approximation to the inverse of the covariance matrix is
\begin{equation}
\begin{split}
\label{riomain}
C^{-1} &\approx n\left[\sum_{j,k=1}^d\lvert\bra{R_k}O_0\ket{R_j}\rvert^2\ket{R_k,R_j}\bra{R_k,R_j}\right. \\
&+ \left.\sum_{j\ne k=1}^d \bra{R_j}O_0\ket{R_j} \bra{R_k}O_0\ket{R_k}\ket{R_j,R_j}\bra{R_k,R_k}\right].
\end{split}
\end{equation}
 Our initial observable, $O_0 =J_z$, anti-commutes with $R$, meaning that
\begin{equation}
RJ_zR^\dagger = -J_z.
\end{equation}
Because of this, we see that 
\begin{equation}
\langle R_j^{(-)}|J_z|R_k^{(-)}\rangle = - \langle R_j^{(-)}|J_z|R_k^{(-)}\rangle = 0,
\text{ for }j,k= 1,\ldots,a,
\end{equation}
and
\begin{equation}
\langle R_j^{(+)}|J_z|R_k^{(+)}\rangle = - \langle R_j^{(+)}|J_z|R_k^{(+)}\rangle = 0,\linebreak \text{ for}j,k= a+1,\ldots,d.
\end{equation}
As discussed above in Eq. \ref{eq:TraceInvC}, we know that after time $t_n$, the trace of the inverse covariance matrix is given by ${\rm Tr}({\mbf C}^{-1}) = n\beta$, where $\beta = \| \mathcal{O}(0) \|^2$ is a constant independent of the Floquet map, $U_{\tau}$, driving the system. 

Thus, the matrix representation of $J_z$ in the ordered basis in which $R$ is diagonal is \emph{anti-block diagonal}. We immediately see that Eq. \ref{riomain} simplifies to
\begin{equation}
\label{invCSimple}
C^{-1} \approx n\left[\sum_{j,k=1}^d\lvert\bra{R_k}O_0\ket{R_j}\rvert^2\ket{R_k,R_j}\bra{R_k,R_j} \right].
\end{equation}
In this basis, $C^{-1}$ is approximately diagonal. So we can actually give an analytical formula for the Shannon entropy. Remember that we previously defined the normalization factor as $\beta$. So the eigenvalues of $C^{-1}$ are simply $\lvert\bra{R_k}O_0\ket{R_j}\rvert^2/\beta$.
To compute the expected value of the Shannon entropy, we use the results of Wootters \cite{Wootters} for the expected value of entropy of the entries of a state expressed in a random basis, and sampled from the appropriate ensemble. We see that since $J_z$ is anti block diagonal, there are only $2\times(d-1)/2\times(d+1)/2$ nonzero terms. 
 
Now, we can directly use Wootters formula for the expected Shannon entropy,
\begin{equation}
H_{exp} = \log(D)-0.729637 = \log\left(\frac{d^2-1}{2}\right)-0.729637.
\end{equation}
For $d = 21$, we get $H_{exp} = 4.66$.  Numerically, we find a somewhat larger value for the kicked top, $H_{KT} = 4.85$ and $H_{av} = 4.69$ for entropy averaged over 100 block diagonal COE matrices. This is due to the fluctuations in $H$ about the expected value and these fluctuations reduce as we increase $d$ and we find an excellent convergence with the analytical expression derived above.

2) \textbf{The  CUE} : The above analysis can be carried over to a quantum map that does not have time reversal symmetry. 
In this case, since there is no parity symmetry, there are   $d^2-1$ nonzero terms in Eq. \ref{riomain}, and therefore, we get for the expected Shannon entropy as
\begin{equation}
H_{exp} = \log\left({d^2-1}\right)-0.729637,
\end{equation}
which agrees remarkably well with our numerical simulations (for $d = 21$, $H_{exp} = 5.35$).

3) \textbf{A different Haar random unitary at each time step}: In this case, we explore the complete Hilbert Space and we get $H_{exp} = \log(d^2-1)$, which agree very well with our simulations (for $d = 21$, $H_{exp} = 6.08$). The maximum possible value of the mutual information is attained when all eigenvalues of the covariance matrix are equal.  In order to extract the maximum information about a random state, we must measure all components of the Bloch vector with maximum precision.  In finite time, we obtain the best estimate by dividing evenly among all observables. 

\section{Conclusion and Outlook}

 The missing information in deterministic chaos is the \textit{initial condition}. A time history of a trajectory at discrete times is an archive of information about the initial conditions given perfect knowledge about the dynamics. Moreover, if the dynamics is chaotic the rate at which we learn information increases due to the rapid Lyapunov divergence of distinguishable trajectories and we expect unbiased information because of the ergodic mixing of phase space.  That is, if the information is generated by chaotic dynamics, the trajectory is random, and all initial conditions are equally likely until we invert the data and discover the initial state.   

 Dynamics sensitive to the initial conditions will reveal more information about the initial conditions as one observes the system trajectory in the course of time. 
  Classically chaotic dynamics generates this unpredictability, or information to be gained about the initial coordinates of the trajectory. 
  Similarly, we found that the rate at which one obtains information about an initially unknown quantum state in quantum tomography is correlated with the extent of chaos in the system. This is a new quantum signature of classical chaos. In fact, our results can be regarded as signatures of chaos in quantum systems undergoing unitary evolution, as measurement backaction is negligible. We have been able to quantify the information gain using the FI associated with estimating the parameters of the unknown quantum state. When the system is fully chaotic, the rate of information gain agrees with the predictions of random matrix theory. 

 At its core, our approach is akin to the Kolmogorov-Sinai (KS) entropy measure of chaos \cite{s59}.  Incomplete information about the initial condition leads to unpredictability of a time history.  In the presence of classical chaos, in order to predict which coarse-grained cell in phase space a trajectory will land at a later time, we require an exponentially increasing fine-grained knowledge of the initial condition.  The KS entropy is the rate of increase, and is related to the positive Lyapunov exponents of the system. Is there a meaningful quantum definition of KS entropy?  Our results seem to suggest this.
 In order to predict the measurement record with a fixed uncertainty, we need to learn more and more about the initial condition.  Is the rate at which we obtain this information exponentially fast when the system is quantum chaotic?  Does this converge to the classical Lyapunov exponents in the  limit of large action (small $\hbar$)? There are many important subtleties in these questions. 
 
  As we gain more and more information, eventually quantum backaction becomes important in the measurement history.  The number of copies we have and the shot noise on the probe limits the ultimate resolution with which we can deduce the quantum state \cite{crd14}.  Unlike classical dynamics, we can never consider infinite resolution, even in principle.  The quantum resolution is limited by the size of $\hbar$. As the dimension of the Hilbert space increases, and hence the effective $\hbar$ decreases, we expect 
 to see an even sharper difference in the information gain as a function of chaoticity. In the limit when $d$, the dimension of the Hilbert space, becomes infinity, we expect the rate of information gain to be intimately related to the classical Lyapunov exponents.
 How all this translates into a quantum definition of KS entropy is an important subject of further investigation. 
 
 In principle we never have perfect knowledge of the dynamics.  This is related to hypersensitivity to perturbations\cite{Scott/Caves} in quantum chaotic dynamics.  This implies that, though quantum systems show no sensitivity to initial conditions, due to unitarity, they do show a sensitivity to parameters in the Hamiltonian~\cite{per00, sc96}.
 How does this fundamentally limit our ability to perform quantum state reconstruction when the system is sufficiently complex, and the equivalent dynamics is chaotic.  This poses interesting questions for quantum tomography and, more interestingly, for quantum simulations. Under what conditions are the system dynamics sensitive to perturbations and how does this effect our ability to perform quantum tomography? Under what conditions does the underlying quantum chaos affect our ability to accomplish quantum simulations in general? We hope to address these questions in our future work.

\acknowledgments
We acknowledge useful discussions with Prof. Shohini Ghose and Prof. Arul Lakshminarayan.
VM acknowledges support from NSERC, Canada (through Discovery grant to M. Doebeli). CR acknowledges the support by the Freie Universität Berlin within the Excellence Initiative of the German Research Foundation. IHD acknowledges support of the US National Science Foundation, Grants, PHY-1212445 and PHY-1307520.

\bibliographystyle{pramana}

\end{document}